\documentclass[11pt]{amsart}
\usepackage{amscd,amsthm,amssymb,amsfonts,amsmath}
\newfont{\frakturfont}{eufm10 scaled\magstep1}

\newcommand{\db}{\bar{\partial}}

\newcommand{\p}{\partial}
\newtheorem{theorem}{Theorem}[section]

\title{Quantum Mechanical Mirror Symmetry, D branes, and B fields}
\author{Mark Stern}
\date{}
\begin{document}

\begin{abstract} We construct quantum mechanical models which mimic many
features of string theory. We use these models to gain improved descriptions of
B fields and gerbes. We examine the analogs of $T$ duality, D branes, and mirror
symmetry in these models and derive quantum mechanical analogs of standard
 phenomena, such as the noncommutative geometry induced by a B field.
 \end{abstract} 
 \maketitle
\section{Introduction}
In this note we attempt to gain insight into mirror symmetry from a quantum
mechanical perspective. Replacing the superconformal field theories 
of superstring theory with quantum mechanics leads to constructions which are 
extremely elementary (and which can be expressed in terms easily accessible to
 a mathematician with no familiarity with physics). Of course, such elementary 
 constructions do not lead to such deep predictions as the enumeration
of rational curves. Nonetheless,  our quantum mechanical analogs 
appear, in many cases, to provide low energy approximations to string theoretic
phenomena. These approximations can then be applied to gain new understanding 
of some poorly understood features of string theory, most notably B fields and 
gerbes.

We begin with an examination of supersymmetric quantum mechanics with 4 real
 supercharges 
on a Calabi Yau manifold, $X$. 
We recall here, for the reader completely unfamiliar with mirror symmetry, some of
 mirror symmetry's coarse features. (See for example \cite{CK}).  
One associates a superconformal field theory to an n complex dimensional 
Calabi Yau manifold $X$. This
field theory generates two filtered 
differential complexes, which might be called the $A'$ and $B'$ complexes. The
$A'$ 
complex can be identified with the de Rham complex on $X$, with a deformed ring
 structure. Let $T_X^{1,0}$ denotes the holomorphic tangent bundle of $X$. The 
 $B'$ 
 complex can be identified with the Dolbeault complex with the ring structure
  obtained by identifying $H^{n-p,q}(X)$ with $H^q(W,\bigwedge^pT_X^{1,0}).$ Here 
  the ring structure on $H^q(W,\bigwedge^pT_X^{1,0})$ is
given by exterior product of forms and tangent vectors. 
If two manifolds are mirror, there is an isomorphism of the associated field
theories which exchanges these two complexes. In particular, when two 
Calabi Yau manifolds $M$ and $W$ of complex dimension $n$ form a mirror pair
 then: 
\begin{enumerate}
\item $h^{p,q}(M) = h^{n-p,q}(W).$
\item  There is a natural local correspondence between the complex moduli of $M$
and the Kahler moduli of $W$.
\item The Hodge filtration on $M$ is mirror to a filtration on $W$ which
converges in the large Kahler limit to a filtration by degree. 
\item The Dolbeault cohomology rings on $M$ and $W$ are isomorphic with 
respect to the quantum deformed ring structure of
$H^{\cdot,\cdot}(M),$ and the ring structure on $W$ obtained via the
isomorphism $H^{n-p,q}(W)\simeq H^q(W,\bigwedge^pT_W^{1,0}).$
\end{enumerate} 
We construct two filtered differential complexes on $X$,
which we will call the $A$ and $B$ complexes. These play the role of the 
$A'$ and $B'$ complexes above. In particular, they are simply de Rham and 
Dolbeault complexes. The $A$ complex has the Hodge filtration as usual. The 
filtration on the Dolbeault complex, seems to be new, however, and has 
interesting features closely parallel to those appearing in mirror
symmetry. In particular, features (1),(2), and (3) above have their analogs in
our construction. The deformations of ring structure that we find seem, however, to lack 
the interesting features of those given by string theory. As an amusing 
consequence of our examination of these complexes, we find a slight
strengthening of the Hodge theorem for compact Kahler manifolds. 

A quantum mechanical analog of mirror symmetry should have spectra of
Hamiltonians preserved under mirror symmetry. The simple example of an
elliptic curve immediately shows that this fails for the supersymmetric quantum 
mechanics discussed in the first half of this paper. In order to remedy this, 
we construct a quantum mechanical analog of T duality and show formally how to
 enlarge both our Hamiltonian 
and the space of sections on which it acts so that, at least for tori, 
T duality preserves the spectrum. We warn the more mathematical reader that 
the construction of these models (sections 7- 10) is carried
out in a formal fashion, ignoring questions of what types of convergence to
impose on various infinite series. 

  The introduction of an analog of T duality immediately leads to analogs of D
  branes and to a very simple explicit realization of B fields, gerbes, and
  n-gerbes. The effects of the B field in our quantum mechanics include the 
  automatic introduction of a Moyal product and a modification of the metric 
to the effective metric seen by open strings in the presence of a B
 field (\cite{SW}). The realization of a gerbe provided by our model
 distinguishes between gerbes appropriate for closed strings and those suitable
 for both open and closed strings. The obstruction to a closed string gerbe 
 being lifted to an open string gerbe is given by the cohomology class of the
 field strength of the $B$ field.
 The extension of this gerbe construction to n gerbes, $n>1$, does not appear
  to be suitable for modelling the effects of the Ramond-Ramond fields. In 
   subsequent work, we will offer a different method for including
    these fields in a quantum mechanical model.

\section{The complexes}\label{ABC} 
Consider a supersymmetric quantum mechanics with 4 real supercharges. Thus we 
have 4 self adjoint operators 
$$Q_1,Q_2,Q_3,Q_4,$$ 
densely defined on some Hilbert space. The operators satisfy the algebra 
\begin{equation}\label{alg}
Q_iQ_j+Q_jQ_i = 2\delta_{ij}H,
\end{equation}
for some Hamiltonian $H$. 

We may realize this algebra by fixing our Hilbert space to be the $L_2$
differential forms on a complete Kahler manifold $X$ and setting  
$$Q_1 = \db + \db^*, iQ_2 = \db - \db^*, Q_3 = \p + \p^*, iQ_4 = \p - \p^*.$$
The Kahler identities guarantee that (\ref{alg}) is satisfied with 
$2H = \Delta$. We could also, of course,  realize the algebra in other ways, for
example, by setting 
$$Q_1'= d+d*, iQ_2' = d - d^*, Q_3' = d_c+d_c^*, iQ_4' = d_c-d_c^*.$$

With this data, we may construct our complexes. 
   The $A-$ complex is simply the de Rham complex, 
$(A^{\cdot},d,F^{\cdot})$, equipped with the Hodge filtration:  
$$F^pA^j = \oplus_{k=0}^{j-p}A^{p+k,j-p-k},$$
for $j\geq p,$ and zero for $p>j$.
 The Hodge filtration induces 
the decomposition $d = \db + \p$ and leads to the associated graded complex 
$(\oplus_p A^{p,\cdot},\db).$ Let $H_A^{\cdot}$ denote the cohomology of the 
$A$ complex, and let $H^{\cdot,\cdot}_A$ denote the cohomology of the 
associated gradeds.  Then we have the usual Hodge decomposition of the
cohomology, 
$$H^k_A = \oplus_{p+q=k}H^{p,q}_A.$$
(Here the cohomology is to be interpreted as the reduced $L_2$ cohomology
 if $X$ is not compact.)

 The B complex is a filtered Dolbeault complex which we denote 
$(B^{\cdot},\db,K^{\cdot}).$ Here 
$$B^p := \oplus_{j=0}^pA^{n-p+j,j}.$$
The filtration $K^{\cdot}$ depends on 
both the metric and the complex structure. We will define it in terms of 
differential operators analogous to $\db $ and $\partial$ in the $A$ complex.

Choose a unit complex number $b$ and define 
$$P := (\db + b\p^*),$$
and 
$$P' := (\db - b\p^*).$$

Then 
$P$ and $P'$ sum to $2\db$, square to zero, and anticommute. In Kahler normal 
coordinates centered at a point $a$ 
$$P = \sum_j(e(d\bar z_j)-be(dz_j)^*)\frac{\p}{\p\bar{z}_j} + O(a),$$
where $e(X)$ denotes exterior multiplication by $X$, $e^*$ denotes the adjoint
operation, and $O(a)$ denotes terms vanishing to first order at $a$.
Similarly, 
$$P' = \sum_j(e(d\bar z_j)+be(dz_j)^*)\frac{\p}{\p\bar{z}_j} + O(a).$$
At the origin of a Kahler normal coordinate system centered at $a$, we define 
$$u^j = e(d\bar z_j)-be(dz_j)^*, v^j = e(d\bar z_j)+be(dz_j)^*, $$
and for a multiindex $I$, set $u^I = u^{i_1}\cdots u^{i_{|I|}},$ and similarly 
define $v^I$. 

The operators $P$ and $P'$, naturally defined on a Kahler manifold with a 
choice of unit complex number $b$, lead us to define summands of 
$B^{\cdot}$ by 
$$B^{p,q} = \{f\in K^{p+q}: f = \sum_{|I|=p,|J|=q}f_{IJ}u^Jv^Idz_1\wedge\cdots\wedge dz_n)\}.$$
Then we have the decomposition, 
$$B^j = \oplus_{p+q=j}B^{p,q}.$$ 
We now define the filtration $K^{\cdot}$ analogously to the Hodge filtration. 
For $j\geq p,$
$$K^pB^j = \oplus_{k=0}^{j-p}B^{p+k,j-p-k},$$ 
and zero otherwise.

 The Kahler identities allow us to extend the usual proof
of the Hodge decomposition to this bigraded complex, giving 
$$H_B^k = \oplus_{p+q=k}H^{p,q}_B,$$
where $H_B^{\cdot}$ denotes the cohomology of the $B$ complex and $H^{p,q}_B$ 
is the cohomology of the graded complex (interpreted as reduced $L_2$ cohomology
if $X$ is not compact). 

\section{Identifying $H_B^{p,q}$}
Observe that a harmonic representative of an element of $H^{p,q}_B$ is also a 
harmonic representative of an element of the de Rham cohomology $H^{\cdot}_A$, 
but need not be of fixed degree with respect to the (usual) degree of the 
$A$ complex. So, do the spaces $H^{p,q}_B$ lead to new invariants of Kahler 
manifolds? We will see below.

Standard arguments show that the complex $(B^{p,\cdot},P)$ is a flabby 
resolution of the kernel of $P$ on $B^{p,0}$. Computing in local coordinates,
we see that this latter sheaf is isomorphic to $\Omega^{n-p}$, the holomorphic
 $n-p$ forms. Hence, we find that 
\begin{equation}\label{mirror1}
H^{p,q}_B(X) \simeq H^{n-p,q}_A(X).
\end{equation}

Note, however, that the harmonic embedding of $H^{p,q}_B(X)$ in 
$H_A^{\cdot}(X)$ does {\em not} lie in $H^{n-p,q}_A(X)$. So, the actual
 embedding may hold new data for us.
   In order to analyze this embedding, we let $L$ denote exterior multiplication 
 by the Kahler form and $L^*$ its adjoint. Observe that the Kahler identities
   $$\partial^* = [\db,-iL^*],$$
   and
    $$[\partial^*,-iL^*] = 0,$$
 imply 
$$e^{ibL^*}\db e^{-ibL^*} = P.$$
Thus the associated graded complex of the $B$ complex is {\em conjugate} to the 
original $B$ complex. This is a rather strange result whose analogous 
statement for the $A$ complex would be that the (global) Dolbeault complex is isomorphic 
(not even just quasi-isomorphic) to the de Rham complex. Assuming that there is a
 strong (mirror) correspondence between the $A$ and $B$ complex, we are led to
  discover the following strong version of the Hodge decomposition theorem. 
\begin{theorem}
Let $X$ be a compact Kahler manifold. 
There exists a bounded pseuodifferential operator $h$ of order zero 
so that 
$$e^{-h}de^{h} = \db.$$
In particular, the de Rham complex (of global sections) is isomorphic to the Dolbeault complex 
via the map from de Rham to Dolbeault:
$$f\rightarrow e^{-h}f.$$
\end{theorem}
\noindent
Proof: The proof is trivial once one suspects the existence of such an $h$. 
$h$ may be chosen (nonuniquely) to be  
$$h := -d^*\p\Delta^{-1}.$$
Here we follow the convention that $\Delta^{-1}$ vanishes on the harmonic 
forms, has range perpendicular to the harmonic forms, and satisfies 
$(dd^*+d^*d)\Delta^{-1} = Id$ on the orthogonal complement to the 
harmonics. $\blacksquare$

Observe that for the $B$ complex we conjugate by a local operator to obtain an
isomorphism between the $\db$ and $P$ complexes but 
could also construct a global operator of a form analogous to $h$ above. 
Clearly, however, there is no local operator conjugating $d$ to $\db$.

\section{Variation of structure}
Now specialize to the case where $X$ has a metrically trivial canonical bundle. 
We will call such an $X$ Calabi-Yau (or CY). This is a slight abuse of notation
as one usually requires the vanishing of $H^{p,0}_A$ for $p<n$. Let $\omega$
denote a generator for $H^{n,0}$. 

We list some correspondences between forms (up to scalars) in the $A$ and $B$
complexes given by relating operators and various harmonic generators of 
one dimensional spaces $H_A^{p,q}$ and $H_B^{p,q}$. 

$$\begin{array}{cc}
\mbox{A complex} & \mbox{B complex}\\
1 & \omega\\
\omega & e^{ibL}\\
\bar\omega & e^{-i\bar b L}\\
dvol & \bar\omega\\
d+d^* & \db+\db^*\\
\db+\db^* & d+bd^*.
\end{array}$$
If $M$ is finite volume, the correspondence between $dvol$ and $\bar\omega$ 
suggests that the bounded linear functional on the $A$ complex obtained by 
integrating a form over $M$ should be paired with the functional on the $B$
complex obtained by taking the $L_2$ inner product with $\bar\omega$.

For a CY 
manifold, $$H_A^{n-p,q}\simeq H^{-p,q},$$
where $H^{-p,q}$ denotes cohomology with coefficients in the p-th exterior 
power of the holomorphic tangent bundle. The isomorphism is given by 
sending $f_{IJ}d\bar{z}^I\theta^J$ to 
$f:=f_{IJ}e(d\bar{z}^I)\theta^J\omega$, the section 
$\theta^J$ of $\bigwedge^pT_X$ acts by interior multiplication. 
In kahler normal coordinates, we write the latter as 
$f_{IJ}d\bar{z}^Ie^*(dz^J)\omega$.  Now an elementary computation shows 
that the map
$$f\rightarrow f_{IJ}u^Iv^J\omega,$$
gives the isomorphism (on harmonic representatives) between 
$H^{n-p,q}_A(X)$ and $H^{p,q}_B(X).$ We can also write the image of $f$ under 
this map as 
$$F_L:f\rightarrow
f_{IJ}e^{ibL^*}e(d\bar{z}^J)e^{-ibL^*}e^{ibL}e^*(dz^I)e^{-ibL}\omega.$$ 

As we noted in the introduction, mirror symmetry is supposed to exchange 
metric data with complex structure data.
Hence the analog of the variation of Hodge structure on the $A$ complex, which
is, of course, obtained by varying the complex structure would be the 
variation of $H^{p,q}_B$ as we vary the metric. Holding the complex structure 
fixed, this corresponds to varying the Kahler form. So, a reasonable
analog of variation of Hodge structure is studying the infinitesimal 
variation of $F_L$ as $L$ varies. This is easiest to do for $H_B^{0,q}$ and 
$H^{p,0}_B$, for in those cases harmonic representatives in $H^{-p,0}$ and 
$H^{0,q}$ are independent of the Kahler class.  
For $f\in H^{-p,0}$ we get 
$$F_L(f) = f_Ie^{ibL}\theta^Ie^{-ibL}\omega
 = f_Ie^{ibL}\theta^I\omega.$$
Hence, the first variation of $F_L(f)$ is 
 
$$ibL_1F_L(f),$$
where $L_1$ denotes exterior multiplication by the first variation
 of the kahler 
form. Similarly, higher variations are easily computed in terms of further wedge
 products. 

Next let us consider the large Kahler structure limit of the filtration
$\bar K^{\cdot}$. For $l\geq p,$ we have 
$$\bar K^pB^l = \{f\in B^l:f = \sum_{|I|=p}u^If_I, f_I\in B^{l-p}\}.$$
The above decomposition, $f = \sum_{|I|=p}u^If_I$, need not be uniquely defined.
For simplicity, we consider a 1 parameter
variation of Kahler structure; i.e., we simply considering scaling, replacing 
$$L\rightarrow TL, T\rightarrow\infty.$$
This replaces $L^*$ by $T^{-1}L^*$. Since we will be varying the metric, we 
choose local Kahler normal coordinates $\{z^i\}$ with respect to $T=1$. Then 
we rewrite $\bar K^pB^l$ as 
$$\bar K^pB^l = \{f\in B^l:f = \sum_{|I|=p}e^{ibL^*/T}e(d\bar{z}^I)e^{-ibL^*/T}f_I, f_I\in B^{l-p}\}.$$ 
In particular, taking a limit as $T\rightarrow \infty,$ we obtain a limiting 
filtration 
$$\bar K^p_{\infty}B^l = \{f\in B^l:f = \sum_{|I|=p}e(d\bar{z}^I)f_I, f_I\in B^{l-p}\}.$$  
This filtration reduces on $B^l$ to the filtration by degree: 
$$\bar K^p_{\infty}B^l = \{f\in B^l: \mbox{degree}f\geq n-l+2p\},$$ 
where degree is in the usual sense of degree of a differential form. 
 
Thus, as with mirror symmetry, we obtain a filtration which is a deformation of
the degree filtration. 

We remark that in the case of $h^{1,1}_A = 1,$ the freedom to choose the phase
 $b$ in our construction effectively makes our construction dependent on the 
 complexified Kahler cone rather than the real Kahler cone. This gives $b$ a
 role similar to that of a $B$ field. This similarity recurs in Section \ref{TDM}.

\section{Products}

 Our analogy with mirror symmetry weakens considerably when we turn to product
  structures. Our constructions 
 provide a natural deformation of the ring structure of the $B$ complex, and not the $A$ complex as with
  mirror symmetry. 
  
We consider ring structures to be determined, in part, by the differentials of
 our complexes; we require the differentials to satisfy a graded Leibniz rule. 
 Thus, before we introduce the filtration $K^{\cdot}$ and the differentials 
 $P,P'$, we have two obvious ring structures on the
  complex $(B^{\cdot},\db)$. (We continue to restrict to the case of CY 
  manifolds). The first ring structure 
is the usual one given by the wedge product. The second is induced, as
 previously mentioned, by writing forms as 
 $g = g_{IJ}d\bar z^J\theta^I\omega$ and $h = h_{LK}d\bar z^K\theta^L\omega,$ 
 taking their product to be 
 $$g_{IJ}h_{LK}(-1)^{|I||K|}d\bar z^{J\cup K}\theta^{I\cup L}\omega.$$
 The operators $P$ and $P'$ are also derivations with respect to the second of
  these ring structures. Moreover, with this structure, 
the map $F_L:H^q(M,\bigwedge^p T_M^{1,0})\rightarrow H^{p,q}_B$ is a ring
isomorphism. Thus our complexes do not seem to introduce any natural
 deformations of ring structure.

  We can use $F_L$ to impose a deformed ring structure if we consider the $B$
 complex equipped with the usual wedge product. Consider the string of
  isomorphisms 
  $$H_A^{n-p,q}\simeq H^q(M,\bigwedge^pT_M^{1,0})\stackrel{F_L}{\simeq}
  H_B^{p,q}.$$
  
  For the $A$ complex it is more natural to express 
  a form $f$ as  $f = \hat f_{KJ}d\bar z^Jdz^K$ rather than 
  $f = f_{IJ}e(d\bar z^J)e^*(dz^I)\omega$. Here (in Kahler normal
  coordinates as always), $f_{KJ} = \pm \hat f_{K^cJ}.$ Then we may write 
$$F_L(f) =
f_{IJ}e^{ibL^*}e(d\bar{z}^J)e^{-ibL^*}e^{ibL}e^*(dz^I)e^{-ibL}\omega$$
$$ =
\hat f_{KJ}e^{ibL^*}e(d\bar{z}^J)e^{-ibL^*}dz^Ke^{ibL},$$ 

Set 
$$\hat F_L(f) = \hat f_{IK}e^{ibL^*}e(d\bar{z}^I)e^{-ibL^*}dz^K.$$
Then imposing a ring structure on the $A$ complex by making 
$\hat F_L$ a ring isomorphism with respect to the wedge product on the image
complex defines a deformation of the product structure on the $A$ complex, which 
in the large $L$ limit obviously converges to the usual ring structure. This product does not appear to be
analogous to the deformation of the ring structure obtained in mirror symmetry, 
but we see that we have deformed
the usual product in such a way that as the metric grows (and 
therefore $L^*$ shrinks) we return to the usual product structure. We are 
viewing forms as operators here. Under this identification, the "vacuum vector" 
i.e., the "unit" degree zero vector in the exterior algebra passes from 
$1$ to $e^{ibL}$. This is also the natural vacuum vector if we 
define a grading of our complex using $d+d_c^*$ instead of $d$.

An alternate approach to constructing a deformed product from this structure is
to consider analogs of the Yukawa coupling:
$$(u_1,u_2,u_3):=
\int_M\omega\wedge\nabla_{u_1}\nabla_{u_2}\nabla_{u_3}\omega,$$
where $\nabla$ denotes the Gauss Manin connection. 
If we consider $\omega$ to be a generator for $H_A^{n,0}$, we have seen that 
the analog in the $B$ complex is 
$$e^{ibL},$$ 
variations of which are simply given by wedge products with appropriate forms. 
This approach seems to return the undeformed wedge product. 

\section{Dropping the Calabi Yau Hypothesis}
If we drop the assumption that our manifold has trivial canonical bundle, the
preceding construction becomes modified in an obvious fashion. The construction
of the $A$ and $B$ complexes is unchanged, but we lose the identification of 
$H_B^{p,q}$ with $H^{-p,q}$ and consequently any ring structure associated to 
this identification. Nonetheless, we do not lose the basic isomorphism, 
$$H^{n-p,q}_A \simeq H^{p,q}_B.$$ The actual map realizing this isomorphism on 
harmonic representatives must be re-expressed in the obvious fashion as 
$$F_L(f_{IJ}d\bar z^J dz^I) = f_{IJ}e^{ibL^*}d\bar z^J e^{-ibL^*} e^{ibL}dz^I.$$

\section{Quantum Mechanical T Duality}\label{QMT}
Let us recall from a geometric perspective, the rules for 
T-duality worked out for WZW sigma models in Buscher \cite{B1} and
\cite{B2}.  
Suppose that we are considering a manifold $M$ which is a Riemannian 
oriented circle bundle over a base manifold $B$. In particular, we assume 
that the associated circle action is an isometry. Let $U_a$ be a coordinate
cover of $B$ and set $V_a = \pi^{-1}(U_a)$, where $\pi:M\rightarrow B$ is the
 projection. Choose coordinates $(x_a^i,x_a^0)$ for $V_a$, where $x_a^0$ 
 is a choice of angular coordinate for the fiber and $x_a^i$ are coordinates
  for the base, $U_a$. Buscher derives the following
transformation rules for the metric, $g$, the $B$ field $b$, and the
dilaton $\phi$. $T-$ duality replaces these quantities with 
$\hat g,\hat b,\hat \phi$, where
$$\hat g_{00} = 1/g^{00},$$
$$\hat g_{0i} = b_{0i}/g^{00},$$ 
$$\hat g_{ij} = g_{ij} - (g_{0i}g_{0j} - b_{0i}b_{0j})/g_{00},$$
$$\hat b_{0i} = -\hat b_{i0} = g_{0i}/g_{00},$$
$$\hat b_{ij} = b_{ij} + (g_{0i}b_{0j} - b_{0i}g_{0j})/g_{00},$$
and
$$\hat \phi = \phi - \frac{1}{2} log(g_{00}).$$

Let us understand these transformations more geometrically. First we assume
 that $b = 0$. The metric $g$ can be rewritten as
$$\hat g_{ij}dx^i\otimes dx^j + g_{00}(dx^0 + A_idx^i)^{\otimes 2},$$
where $A_i = g_{0i}/g_{00},$ is independent of $x^0$ since we the circle action
is an isometry.
Thus the horizontal lift of the vector $\frac{\partial}{\partial x^i}$ is the 
vector 
$$H_i = \frac{\partial}{\partial x^i} - A_i\frac{\partial}{\partial x^0}.$$
Observe that 
$$<H_i,H_j> = \hat g_{ij}.$$
Because the $H_i$ are the horizontal lifts of the vectors 
$\frac{\partial}{\partial x^i}$ on $U_a$, they transform as these vectors do 
under a change of coordinate neighborhoods. Hence, we see that the metric 
$\hat g$ is realized (as is well known) on the trivial bundle 
$B\times S^1$.  
We would like to understand the role of the B field better. At first
approximation it is supposed to define a potential for the torsion tensor, 
$T$ defined by 
\begin{equation}\label{tors}
\hat\nabla_{\frac{\partial}{\partial x^a}} \frac{\partial}{\partial x^b}
 - \hat\nabla_{\frac{\partial}{\partial x^b}} \frac{\partial}{\partial x^a} = 
 T_{ab}^c \frac{\partial}{\partial x^b}.
 \end{equation}
 The $T$ dual connection $\hat\nabla$ is assumed to be metric compatible. The torsion is
 related to the $B$ field by 
 $$(db)_{abc} = T_{abc}.$$
 Here we lower indices with the metric $\hat g$.
Observe 
$$\nabla_{H_i}H_j - \nabla_{H_j}H_i
 = (db)_{ij0}\frac{\partial}{\partial x^0}
  = T_{ij}^0\hat g_{00}\frac{\partial}{\partial x^0} =
  T_{ij}^0|dx^0+A_idx^i|^2\frac{\partial}{\partial x^0}.$$

 The similarity of this expression to (\ref{tors}) suggests the possibilty of
  intepreting  $T$ duality literally as an exchange of  
$dx^0+A_idx^i$ and $g^{00}\frac{\partial}{\partial x^0}$; i.e., an exchange of a
vector with its metric dual covector.  In order to make sense
of such an exchange, we need to represent cotangent vectors in some fashion 
akin to differential operators. A natural guess for a representation is
to define a representation on some subspace of partial differential operators
defined by  
$$du^kf(u)\frac{\partial}{\partial u^j} = -\delta^k_if(u)/2\pi,$$
and extend as a ($C^{\infty}(M)$ linear) derivation. 
We will denote this operator 
$$ad(u^k).$$

We have not resolved the question of which subspaces of operators to consider.
In particular, we will in some cases consider formal power series of operators
without addressing in what sense we require convergence. 
As will become clear in later sections, different physical models will require
different subspaces. 

If we want to consider quantum mechanical theories which are invariant under T
duality, then we need to include the T dual of the Laplace operator. 
This should take the form, 
$$g_{ab}ad(x^a)ad(x^b).$$
Thus the dual of the Laplacian is essentially the metric.

Let us consider this construction on a circle of length $2\pi L$.  Write 
$g = L^2 du^2$, with $u$ a $2\pi$ periodic coordinate.  
The formal eigenvectors of $L^2 ad(u)^2$ are then 
of the form 
$$F(u)e^{2\pi p\frac{\partial}{\partial u}},$$
with eigenvector $4L^2p^2$. 
We may interpret this latter operator as translation by $2\pi p$ followed by
multiplication by $F(u)$. This interpretation requires viewing translation by
$p$ as the endpoint of a 1 parameter family of translations. We see that we have a
continuous spectrum for this dual Laplacian. We have, however, forgotten to
 impose a condition. We were intially considering functions which were
 $2\pi$ periodic. In particular, they were annihilated by the operator 
 $$e^{2\pi\frac{\partial}{\partial u}} - 1.$$
 The $T$ dual condition, then, is to restrict to the kernel of 
 \begin{equation}\label{Tcondition}
 e^{i ad(u)} - 1.
 \end{equation}
  Imposing this condition then restricts us to (the span of) 
  those eigen-operators, $e^{2\pi p\frac{\partial}{\partial u}},$
   with $p$ an integer, 
  and we get a spectrum which is fixed under $T$ duality and includes an analog
  of winding modes. 
   In (\ref{BC}) we will give a geometric interpretation of the condition
   (\ref{Tcondition}) in terms of $D$ branes.

 Next, the guess for the role of the $B$ field, based on our $T$ duality
computations is that it enters in a manner similar to a connection term,
 modifying our self dual Laplacian to something
like
$$-\frac{1}{\sqrt[2]{g}}(\frac{\partial}{\partial x^i} + b_{ik}ad(x^k)/i)g^{ij}\sqrt[2]{g}
(\frac{\partial}{\partial x^j}+b_{jk}ad(x^k)/i)
+ g_{ij}ad(x^i)ad(x^j).$$ 
Here we have placed the $B$ field on the Laplacian side, as seems natural for 
a map from $TM:\rightarrow T^*M.$ If instead we collect the $B$ terms on the 
metric side we get (for say constant $g^{ij}$,$b_{ik}$, and $g$)

$$-g^{ij}\frac{\partial}{\partial x^i}\frac{\partial}{\partial x^j}
+ 2ig^{ij}\frac{\partial}{\partial x^i}b_{jk}ad(x^k)
+ (g_{ij}-bg^{-1}b_{ij})ad(x^i)ad(x^j).$$ 
Let $G_{ij}:= g_{ij}-bg^{-1}b_{ij},$ and following (\cite{SW})
(up to a constant) set $\theta^{pi} = ig^{pq}b_{qn}G^{ni}.$ 
Then our operator reduces to 

$$-G^{ij}\frac{\partial}{\partial x^i}\frac{\partial}{\partial x^j}
+ G_{ij}(dx^i+\theta^{pi}\frac{\partial}{\partial x_p})
(dx^j+\theta^{qj}\frac{\partial}{\partial x_q}).$$
In particular, we see that we may interpret this as defining a new metric 
$G$. This is the effective metric seen by open strings in the presence of a B
 field (\cite{SW}).

We next observe that the Moyal product associated to a constant B field 
naturally arises from our formalism. (See \cite{SW} 
and references therein). We initially considered functions and scalar valued 
forms, but were led by considerations of $T$ duality to enlarge our
 coefficients
to include partial differential operators.  We can recover our original space 
of sections as the intersection of the kernels of $ad(x^k)$ for 
all $k$. In the context of open strings, as we shall explain in a subsequent
section, this subspace is analogous to Neumann
boundary conditions. As the B field has the effect of deforming 
\begin{equation}\label{def1}
g_{ij}ad(x^i)ad(x^j)\rightarrow G_{ij}(dx^i+\theta^{pi}\frac{\partial}{\partial x_p})
(dx^j+\theta^{qj}\frac{\partial}{\partial x_q}),
\end{equation}
it is natural to replace the above analog of the Neumann condition by the
condition that we consider sections in the kernel of 
$(dx^i+\theta^{pi}\frac{\partial}{\partial x_p}).$

Formally, we may write a section in the kernel of the deformed $ad(x^i)'s$ 
as 
$$\hat f:= f(x^j + \theta^{jk}\frac{\partial}{\partial x^k}),$$
interpreted as a formal power series in the symbol 
$\frac{\partial}{\partial x^p}.$ Such a section is uniquely determined by 
the zero order term, $f(x)$, in the power series expansion and may be viewed as a 
"flat" lift of the function $f$  to a formal partial differential operator. See 
\cite{Fed}, Chapter 5, for similar constructions in the context of deformation
quantization. The product as differential operators, $\hat f\circ\hat h$  of
 two such flat lifts is 
again a flat lift (by the Leibniz rule) whose zero order term 
is given by $f\ast h$, where $\ast$ denotes the Moyal product with respect to 
 $\theta^{ij}$: 
$$f\ast h = \sum_{k=0}^{\infty}\sum_{|I| = |J| = k}\theta^{IJ}f_Ih_J/k!.$$
Here $\theta^{IJ} := \theta^{i_1j_1}\cdots \theta^{i_{|I|}j_{|I|}},$ 
and $f_I$, for $|I| = k$ and
 denotes $\frac{\partial^k}{\partial x^{i_1}\cdots \partial x^{i_k}}f.$

Thus, we are naturally led to a Moyal product induced by the $B$ field.

\section{Tori}
In this section, as an elementary exercise, we consider the equality 
of the spectrum of 
the operators $\Delta + g_{ij}ad(x^i)ad(x^j)$ on a pair of $T$ dual tori $E_1$
and $E_2$. The
first torus, $E_1$  has area $A$, $B$ field equal to $0$, and complex structure
defined in the usual way by the parameter $\tau = \tau_1 + i\tau_2$. The 
T dual torus has area $\tau_2$, $B$ field given by $\tau_1 ds\wedge dt$, and
 complex structure parameter $\tau = iA$. Thus
 
$$E_1:= C/\lambda Z + \lambda(\tau_1+i\tau_2) Z.$$
We parametrize $E_1$ by 
$$(x(s,t),y(s,t)) = (\lambda s + \lambda \tau_1 t,\lambda \tau_2 t).$$
Set $A = \lambda^2\tau_2$, the area of $T$. Then 
$$dx^2 + dy^2 = \frac{A}{\tau_2}(ds+\tau_1dt)^2 + A\tau_2dt^2,$$
and
$$\Delta = -\frac{\tau_2}{A}\frac{\partial^2}{\partial s^2}
 - (A\tau_2)^{-1}(\frac{\partial}{\partial t} - \tau_1\frac{\partial}{\partial s})^2.
$$
According to our definition of the spectrum for the "metric",
we find the spectrum of $\Delta + dx^2+dy^2$ is given by 
$$\frac{A}{\tau_2}(m+\tau_1n)^2 + A\tau_2n^2 + \frac{\tau_2}{A}p^2
 + (A\tau_2)^{-1}(q - \tau_1p)^2,$$
 with $m,n,p,q$ integer.

Now a T duality transformation changes the metric to 
$$\frac{\tau_2}{A}ds^2 + A\tau_2dt^2,$$
and the Laplacian becomes 
$$-\frac{A}{\tau_2}\frac{\partial^2}{\partial s^2}
 - (A\tau_2)^{-1}\frac{\partial^2}{\partial t^2}.$$ 

According to our earlier discussion, the duality transformation should also 
introduce a $B$ field $\tau_1 ds\wedge dt$, which enters by deforming 
$dx^2+dy^2 + \Delta$ to 

$$\frac{\tau_2}{A}ds^2 + A\tau_2dt^2 + 
-\frac{A}{\tau_2}(\frac{\partial}{\partial s} + \tau_1 dt/i)^2
 - (A\tau_2)^{-1}(\frac{\partial}{\partial t} - \tau_1 ds/i)^2.$$ 

This new operator has spectrum 

$$\frac{\tau_2}{A}p^2 + A\tau_2n^2 + \frac{A}{\tau_2}(m+\tau_1 n)^2
 + (A\tau_2)^{-1}(q-\tau_1 p)^2,$$
 with integer $m,n,p,q$.
 The spectrum is thus invariant under T duality.

\section{D branes}  
\subsection{Dictionary}\label{dict}
In this subsection we construct a dictionary relating operators we have defined 
to standard super conformal field theory constructions. This will allow 
us to interpret some of our results crudely in terms of $D$ branes and permit
the reader to extend these interpretations in many directions.  
Let $X^i(t,s)$ denote, as usual, the pull back to the string world sheet of a
 local coordinate on $M$. We use coordinates $s$ and $t$ on the string,
 satisfying $s$ is constant on the boundary of the string. The translation of 
 stringy constructions into our model begins with the correspondence 
$$X^i_t\rightarrow g^{ij}\nabla_{\frac{\partial}{\partial x^j}},$$
$$X^i_s\rightarrow ad(x^i).$$ 
 
Hence, 
$$\partial X^k \rightarrow g^{kj}\nabla_{\frac{\partial}{\partial x^j}} +
iad(x^k).$$
 
Following the notation of \cite{KO}, we let $Q^{\pm}$ and $J$ denote the left
 moving supercurrents and the $U(1)$ current respectively. 
Then we associate 
$$Q^{+}\rightarrow
 i(e(dz^j) + e^*(d\bar z^j))(\nabla_{\frac{\partial}{\partial z^j}} 
  - ig_{j\bar k}ad(\bar z^k))\sim \partial - \db^*,$$
 
$$Q^{-}\rightarrow 
i(e(d\bar z^j) + e^*(dz^j))(\nabla_{\frac{\partial}{\partial\bar z^j}}
 - ig_{\bar{j} k}ad(z^k)) \sim \db - \partial^*,$$
 
$$\bar Q^{+}\rightarrow
 i(e(dz^j) - e^*(d\bar z^j))(\nabla_{\frac{\partial}{\partial z^j}}
 + ig_{j\bar k}ad(\bar z^k)) \sim \partial+\db^*,$$
 
$$\bar Q^{-}\rightarrow 
-i(e(d\bar z^j) - e^*(dz^j))(\nabla_{\frac{\partial}{\partial \bar z^j}}
 + ig_{\bar{j}k}ad(z^k)) \sim \db+\partial^*,$$ 
 
$$J\rightarrow -ig_{j\bar k}(-e(dz^j)e(d\bar z^k) + e(dz^k)^*e(d\bar z^j)^*) 
+ 2q-2p,$$

$$\bar J\rightarrow ig_{j\bar k}(-e(dz^j)e(d\bar z^k) + e(dz^k)^*e(d\bar
z^j)^*) + 2p-2q.$$

\subsection{Boundary conditions}\label{BC}
In the absence of $D$ branes, the usual boundary condition on an open string is 
the Neumann boundary condition :
$$X_s^i = 0,$$
on the string boundary for all $i$. According to our dictionary, this
corresponds to the condition that 
$$ad(x^i) = 0.$$
Thus the space of scalar valued differential forms is the subspace 
of sections corresponding to standard boundary conditions. Pursuing the analogy,
then a $p$ brane boundary condition corresponds to (locally) forcing the
condition on our sections that they are annihilated by 
$\nabla_{\frac{\partial}{\partial x^i}}$ for $n-p$ directions $p<i\leq n$. The
 sections are
allowed to have coefficients in partial differential operators which involve 
only these same $n-p$ "normal" derivatives. Thus, this 
subspace should correspond to $D$ branes given by $x^i = constant_i$, for 
$p<i\leq n$. (See \cite{P}[p.265-267]). 
We interpret the normal derivatives as determining the position of the
$D$ brane. In the example of the circle considered in Section \ref{QMT}, we saw that the
eigenstates for the metric component of the Hamiltonian were of the form 
$e^{2\pi p\frac{\partial}{\partial u}}.$ This operator is naturally interpreted
as translation by $2\pi p$. Hence, if we view this as determining the relative
positions of the $D$ branes, we find we have 1 brane located at each point of
$2\pi Z$ on the real line if we impose (\ref{Tcondition}). This is the standard
representation for a single brane on a circle. If we do not impose
(\ref{Tcondition}), then the continuous spectrum could be interpreted as a 
family of branes filling space. 

The construction of $D$ branes as in \cite{OOY}, in terms of an equality 
\begin{equation}\label{Oz}
\partial X^m = R^m_n\db X^n,
\end{equation} 
for some orthogonal matrix $R$ generalizes this interpretation of $D$
 branes as choices of subspaces (or possibly quotient spaces) of partial
  differential operators and leads to analogs of $A$ and $B$ branes in 
  our model. We will not pursue this here. 

The notion of D branes being given as kernels of operators does not seem nicely 
self dual. Once we include kernels, one should expect to see also cokernels. 
The arrival of complexes does not seem far behind. It seems likely that our
construction will lead to relations between D branes and holonomic D-modules and
that this will clarify the role of complexes. We have not yet explored this 
possibility.

\subsection{D branes in the presence of a Wilson line}
We next consider an extension of our construction to include the presence of a
background vector bundle. Thus we consider partial differential operators 
acting on sections of a flat $C^n$ bundle $E$. Our partial differential
 operator coefficients are now $end(E)$ valued. Neumann boundary
conditions still are represented by considering the subspace given by the 
kernel of the $ad(x^i)'s$. Suppose we now $T$ 
dualize on a circle factor, in the $x^1$ direction. Assume for simplicty
 that we are in flat space. Then the condition that 
$ad(x^1)$ annihilate our partial differential operator (henceforth $pdo$)
 is replaced by the condition that  
$ad(\nabla_{\frac{\partial}{\partial x^1}})$ annihilate it. Suppose that 
we can choose a frame so that 
$$\nabla_{\frac{\partial}{\partial x^1}} = \frac{\partial}{\partial x^1} + \Lambda,$$
for some constant diagonal $n$ by $n$ matrix $\Lambda$ with eigenvalues
$\lambda_j$. (We are considering a situation like that discussed
in \cite{P}[p.263-267].) The pdos we are allowed then are of the form 
$$f_k(x)(\frac{\partial}{\partial x^1})^k,$$ 
where 
$f_k$ is annihilated by $ad(\nabla_{\frac{\partial}{\partial x^1}}).$  

Expressing $f_k$ as a matrix with respect to a basis in which $\Lambda$ is 
diagonal, we find that the diagonal entries $f_k^{ii}$ are constant in $x^1
$. 
The off diagonal entries $f_k^{ij}$ are of the form an $x^1$ independent
function multiplied by $e^{-x^1(\lambda_i -\lambda_j)}.$ For generic $\Lambda$ 
this is not periodic; therefore $f_k$ is forced to be diagonal, except when two
or more eigenvalues coincide (mod $2\pi iZ)$ at which point there is a gauge
enhancement and the $f_k$ take values in a larger algebra. In order to localize 
the branes as in \cite{P}[p.263-267], we must consider again the $T$ dual of the
periodicity condition, 
$$e^{2\pi\frac{\partial}{\partial u}} - 1.$$
We rewrite this as 
$$e^{2\pi(\nabla_{\frac{\partial}{\partial u}}-\Lambda)} - 1 = 0.$$
Dualizing, this becomes 
$$e^{i ad(u)} - e^{2\pi\Lambda} = 0.$$
Restricting to eigenstates of the metric again, we see that we may write 
$i^{th}$ diagonal entry of our allowed pdos as 
$$F_p^{ii}e^{2\pi(p+\lambda_i)\frac{\partial}{\partial x^1}}.$$
 As in the preceding section, we may
 interpret this as a brane located at position $\lambda_i$ on the circle. 
Hence,  this model reproduces
the expected result that, in the presence of a flat $C^n$ bundle,
 the $T$ dual (along a circle) of the Neumann boundary
condition is $n$ D branes, whose relative position is determined by the
monodromy of the connection. 
This description fails when two or more eigenvalues
coincide, at which point there is an enlargement of the space of sections. 

\section{T duality and Mirror symmetry}\label{TDM}

In this section, I consider the natural question: can we map the A complex to 
the B complex via T duality? 
 There are at least two methods for doing this. The first, which we shall not
 explore here, requires assuming a torus fibration of our manifold (as in
  \cite{SYZ}) on which we may $T$ dualize. For this to succeed, however, we are 
  required to modify our operators somewhat, for example, replacing $\db$ by
   $Q^{-} - \bar Q^{-}$ 
  and $\partial$ by $Q^{+} + \bar Q^{+}$.
  This approach closely tracks the usual SCFT analysis.
The method which we shall instead explore here is to introduce a complex 
version of
 T duality which may be viewed as a restriction of $T$ duality to objects
 associated to the holomorphic tangent bundle and its dual. This approach leads
 to constructions which are weakly reminiscent of $A$ branes and $B$
  branes. 

So, we consider an antilinear $T$ duality replacing 
$\frac{\partial}{\partial z^i}$ by 
$ig_{\bar i j}ad(z^j)$, $e(dz^i)$ by $e^*(dz^i)$, leaving the conjugate variable
 unchanged, 
 and extend by making $T^2 = 1$. Now consider our $A$ and $B$ complexes, with
  their coboundary operators.  
How can we use complex $T$ duality to pass from one to the other? 
Start with 
$$d = 
e(d\bar z^i)\frac{\partial}{\partial \bar z^i}
 + e(dz^i)\frac{\partial}{\partial z^i}.$$
If we $T$ dualize following the above recipe, this becomes 

$$d_T:=e(d\bar z^i)\frac{\partial}{\partial \bar z^i} + ie(dz^i)^*g_{\bar i j}ad(z^j).$$

Now our coboundary operator for the B complex has the form
 (in nice coordinates)
$$P = (e(d\bar z^i)-be(dz^i)^*)\frac{\partial}{\partial \bar z^i}.$$
How can we connect these? We alter our "boundary conditions". Initially
we considered scalar valued forms for our $A$ and $B$ complexes, but in order to
define $T$ duality, we were forced to consider forms with
coefficients in partial differential operators. Our original space is recovered 
by imposing the analog of Neumann boundary conditions: $ad(x^i) = 0$. 
In general, we should expect to impose $dim_R(M)$ conditions to get a complex
"as large as" the original one. (Presumably there is a holonomic condition that
is the correct generalization). The usual complex satisfies 
$ad(z^i) = 0$ and $ad(\bar z^i) = 0$, for all $i$. Let's preserve this as much
 as possible. 
So, we assume for both complexes that $ad(\bar z^i) = 0$. In order for T duality
to transform $d$ to $P$, we must deform the first condition, $ad(\bar z^i) = 0$,
 to 
$ad(z^j) - ibg^{j\bar k}\frac{\partial}{\partial \bar z^k} = 0$. 
On this restricted subspace, $d_T = P$. Note the
similarity between the modification of the condition that $ad(z^i) = 0$ 
and the deformation of the Neumann boundary
condition induced by a $B$ field in (\ref{def1}). This parallel inclines us to
interpret $-ib\omega$ as a $B$ field, except that here we are only using it to
deform $ad(z^i)$ and not $ad(\bar z^i)$.

Thus we consider sections of the form 
$f(z,\frac{\partial}{\partial z^j}+ig_{j\bar k}\bar z^k/b),$ 
again interpreted as formal power series.
These are determined uniquely by $f(z,\frac{\partial}{\partial z})$, 
a collection of holomorphic sections of powers of the holomorphic tangent
 bundle. On the other hand, if we now work backwards, the conditions that 
$ad(\bar z^i) = 0$ and
 $ad(z^j) - ibg^{j\bar k}\frac{\partial}{\partial \bar z^k} = 0$ are $T$ 
dual to the condition (for the $A$ complex) that 
$ad(\bar z^i) = 0$ and 
$\frac{\partial}{\partial z^i} - ib\frac{\partial}{\partial \bar z^i} = 0$. 
This latter condition is, of course, not coordinate independent. Because we 
have a sum of a complex linear and a complex anti-linear map, it is not well
 defined independent of a choice of complex basis. Its definition requires a 
 choice of underlying real structure for the
holomorphic tangent space $T^{1,0} = T_R^{1,0}\otimes C,$ where (at a point)
$T^{1,0}_R$ is a subspace spanned by the $\frac{\partial}{\partial x^i}$
 in some local coordinate system with $z^i = x^i+iy^i$. 
 
Assuming the existence of such a structure, the $A$ complex, for say $b = 1$
consists of  sections of the form 
$$f(y,\frac{\partial}{\partial z}).$$ 
As $b$ varies, these families rotate in an obvious fashion.
Thus, the sections are determined by their values on Lagrangian submanifolds
determined by this real structure and the choice of $b$. 
 
\section{Global Structure of the B field}\label{geomB}
\subsection{Global structures associated to B fields}
We have seen how a $B$ field can alter differential operators and lead to simple 
noncommutative geometries. Now we will examine the global structures associated
to $B$ fields.

We adopt the notation $A\diamond ad(x^j)$ to denote the operator
which sends $(\frac{\partial}{\partial x^j})^k$ to 
$\sum_{p=0}^{k-1}(\frac{\partial}{\partial x^j})^pA(\frac{\partial}{\partial
x^j})^{k-p-1}$. In particular, for pdos of degree greater than one, 
$A\diamond ad(x^j)$ does not denote $ad(x^j)$ followed by multiplication on the
left by $A$. This definition is natural if we wish to associate to a 
$1$ form $b_jdx^j$ a derivation on the space of pdos which is independent 
of coordinate choice.  On the other hand, this operator now
distinguishes between 
$\frac{\partial}{\partial x^i}\frac{\partial}{\partial x^j}$ and 
$\frac{\partial}{\partial x^j}\frac{\partial}{\partial x^i}$. In order to remedy
this, we must consider the space of differential operators with their ordering
 retained; i.e. we do not quotient out by the relations which hold for
 operators acting on the smooth functions. In fact, all of the constructions in
 this section may be defined on spaces of pdos of degree 1, in which case we may
 ignore these ordering difficulties. 
 Hence, for the remainder of this section,  we restrict to that case for
  simplicity. The extension to spaces of higher order pdos is immediate. 
  
 We recall that the $B$ field has the effect of replacing the derivative 
 $\nabla_{\frac{\partial}{\partial x^i}}$ by
   $\nabla_{\frac{\partial}{\partial x^i}} - b_{ij}ad(x^j).$
If $b$ were constant, this shift could be interpreted as conjugation 
$$\nabla_{\frac{\partial}{\partial x^i}}\rightarrow 
e^{(x^ib_{ij})\diamond ad(x^j)}\nabla_{\frac{\partial}{\partial
x^i}}e^{-(x^ib_{ij})\diamond ad(x^j)}.$$ 

This conjugation corresponds to replacing a partial differential operator 
$a_I\partial^I$ by $e^{b_jad(x^j)}a_I\partial^I,$
where $b_j = x^ib_{ij},$ in the example at hand.

If we assume our pdos act on sections of a $U(1)$ bundle, then under a change of
frame, the pdos are modified by $e^{u_j\diamond ad(x^j)}$, where 
$u_jdx^j = g^{-1}dg$ for some $S^1$ valued function $g$. This transforms
$\frac{\partial}{\partial x^i}$ to 
$\frac{\partial}{\partial x^i} + u_i$, $u_i$ acting by the adjoint
action. Here we are using the equality 
$$ad(u_i) = u_{i,j}\diamond ad(x^j).$$

Note that conjugation by the exponential of an arbitrary 1 form,
$$\nabla_{\frac{\partial}{\partial x^i}}\rightarrow 
e^{\lambda_{j}\diamond ad(x^j)}\nabla_{\frac{\partial}{\partial
x^i}}e^{-\lambda_{j}\diamond ad(x^j)},$$ 
will modify $\nabla_{\frac{\partial}{\partial x^i}}$ by (in geodesic
coordinates)
$-\lambda_{j,i}\diamond dx^j,$ with $\lambda_{j,i}dx^i\otimes dx^j$ not
 necessarily a 2 form;
i.e., $\lambda_{i,j}$ need not be skew in its indices. In order for this to be
 a 2 form, it would be necessary that
 $\lambda^i\frac{\partial}{\partial x^i}$ be a Killing vector. On the other
  hand, as noted above, 
 $$ad(\lambda_i) - \lambda_{i,j}ad(x^j) = 0.$$
 Hence, we see that we can interpet this conjugation as transforming 
  $$B\rightarrow B+d\lambda$$
   if we simultaneously transform 
 $$\nabla_i\rightarrow \nabla_i - \lambda_i.$$
 Thus, up to a sign convention, the connection transforms as expected \cite{P}[(8.7.7) and (8.78)]
  under a change of gauge for the $B$ field.

We see that we can interpret $b$ as a component of a connection form for a
 covariant derivative on a space of pdos with allowed transition functions
  of the form $e^{w_j\diamond ad(x^j)}$. It is easy to pass from this 
  perspective to the Deligne cohomology class of a gerbe. 
  Fix an open cover $\{U_{\alpha}\}$ of $M$. Consider collections $p_{\alpha}$ 
   of partial differential operators on each $U_{\alpha}$ such
  that on $U_{\alpha}\cap U_{\beta}$, we have 
  $$p_{\alpha} = e^{w_j^{\alpha \beta}\diamond ad(x^j)}p_{\beta},$$
  for some collection of 1 forms $w_j^{\alpha \beta} dx^j$, satisfying the
  cocycle condition 
  $$w_j^{\alpha \beta} dx^j + w_j^{\beta \mu} dx^j + w_j^{\mu\alpha}
  dx^j = d\lambda^{\alpha\beta\mu}.$$
  
  Here we should either choose 
  $$d\lambda^{\alpha\beta\mu} = 0,$$
  or 
  $$d\lambda^{\alpha\beta\mu} = g_{\alpha\beta\mu}^{-1}dg_{\alpha\beta\mu},$$
  for some $S^1$ valued cocycle $g_{\alpha\beta\mu}$.  
The first choice is the standard cocycle condition for transition functions of
the above form. Note, however, that forcing $w$ to satisfy  $\delta w = 0$ 
(with $\delta$ the Cech coboundary) implies that the field strength 
$$H:=dB$$ 
is cohomologically trivial. 

We may realize the less restrictive condition $\delta w = d\lambda$ for 
nonzero $d\lambda$ if we modify our construction to consider connections not 
on pdos but on gauge equivalence classes of pdos. The action of
 $ad(\lambda_i)$ is trivial on gauge equivalence
classes, and therefore the background gauge field can be dropped from the theory.
One does not expect coupling of the $B$ field to a connection in a closed string
theory; hence, the gauge equivalence class representation should correspond to a
closed string theory, while the representation with $\delta w = 0$ should 
correspond to an open string theory, where one expects a coupling to a gauge
field. In our model then, we see a cohomological obstruction to lifting a $B$
field background from closed to open string theory. The field strength $H$ 
must be cohomologically trivial for a closed string $B$ field to lift to an open
string theory.

Let us consider what data we have now assembled in order to view 
the $B$ field as a connection on the space of pdos (or gauge equivalence
classes of pdos). On each open set $U_{\alpha}$ we have a 2 form $b^{\alpha}$, 
on the intersection of two open sets $U_{\alpha} \cap U_{\beta}$, 
we have a 1 form $w^{\alpha\beta}$, and on each triple intersection,
$U_{\alpha}\cap U_{\beta}\cap U_{\mu}$, we have 
an $S^1$ valued function $g_{\alpha\beta\mu}.$ Let $\delta$ denote the 
Cech coboundary operator. Then this collection satisfies 
$$\delta b = d w, \mbox{   }\delta w = g^{-1}dg, \mbox{  and  } \delta g = 0.$$

Thus, our $B$ field, as numerous authors have predicted (see for example 
 \cite{FW} and \cite{KO}), defines a Deligne cohomology class 
$$(g,w,b)\in H^2(M,D^2),$$
in the notation of \cite{CJM}. 
  
      As $e^{b_j \diamond ad(x^j)}\partial_i = \partial_i + b_i,$ 
 we may interpret this change of gauge as 
a change of connection.  We note that it is the shift that
  follows from tensoring a bundle with a line bundle with connection 
  $\nabla = d + b.$  
  Hence, we may interpret the transition functions associated to our $B$ field 
  construction as corresponding to 
a line bundle with connection on the intersection of each coordinate patch,
 as in Hitchin's picture of gerbes (see \cite{H}).

We compute the curvature of the connection 
$$\nabla_{\frac{\partial}{\partial x^i}} - b_{ij}ad(x^j).$$
Write $\nabla_{\frac{\partial}{\partial x^i}} = \frac{\partial}{\partial x^i}
 + A_i$ in a local frame. Then the curvature is given by 
 $$dx^i\wedge dx^k[\frac{\partial}{\partial x^i} + A_i - b_{ij}ad(x^j),
\frac{\partial}{\partial x^k} + A_k - b_{kp}ad(x^p)] = $$
$$dx^i\wedge dx^k[A_{k,i} - A_{i,k} - (b_{kp,i}+ b_{pi,k})ad(x^p)] =$$
$$dx^i\wedge dx^k[A_{k,i} - A_{i,k} + b_{ik,p}ad(x^p) - db_{ikp}ad(x^p)].$$ 
Using 
$$b_{ik,p}ad(x^p) = ad(b_{ik}),$$

and setting 
$$\tilde F = dA + b,$$
we see that the curvature of our connection can be written 
$$\tilde F - i_p(db)ad(x^p).$$   

Here $i_p$ denotes interior multiplication by $\frac{\partial}{\partial x^p}$. 
Note that here $(dA + b)_{ik}$ acts on a space of partial differential operators by 
the adjoint action. Observe that the combination $\tilde F = dA + b$ is
invariant under both the gauge symmetries of the underlying connection and the
connection on the space of pdos.

\subsection{Higher order gerbes}\label{failsC}
The construction of $B$ fields as connections on spaces of 
pdos can now be iterated. For example, we get a three form connection 
if we consider connections on the $B$ field connections. 

A general $B$ field connection locally has the form 
$$\nabla_i = \partial_i + \lambda_i - b_{ik}\diamond ad(x^k).$$
We define a new $ad$ operator $ad_2$ which acts on these connections by 
$$ad_2(x^i)(\nabla_k) = -\delta^i_k.$$
Then the arbitrary change of $\lambda_i$ and $b_{ik}$ is given by 
$$e^{(P_j + L_{jk}\diamond ad(x^k))\diamond_2ad_2(x^j)},$$
with $L$ a 2 form and $\diamond_2$ the obvious analog of $\diamond$. 
We will view such transformations as a change of frame on our space of $B$ field
connections. 

A connection on the space of $B$ field connections has the local form 
$$\nabla_i^B = \partial_i + \lambda_i + b_{ik}\diamond ad(x^k)
 - W_{ij}\diamond_2ad_2(x^j) - c_{ijk}\diamond
ad(x^j)\diamond_2 ad(x^k).$$
Here $\lambda_i$ and $b_{ik}\diamond ad(x^k)$ act by the adjoint action. 
A change of frame $e^{(P_j + L_{jk}\diamond ad(x^k))\diamond_2ad_2(x^j)}$ now induces 
a change of gauge 
$$\nabla_i^B\rightarrow \partial_i + \lambda_i + b_{ik}\diamond ad(x^k)
- W_{ij}\diamond_2ad_2(x^j) - c_{ijk}\diamond
ad(x^j)\diamond_2 ad_2(x^k)$$
$$ - (P_{j,i} + L_{jk,i}\diamond ad(x^k))\diamond_2ad_2(x^j) = $$
$$\partial_i + \lambda_i - P_i + (b-L)_{ik}\diamond ad(x^k) 
- (W-dP)_{ij}\diamond_2ad_2(x^j) - (c+dL)_{ijk}\diamond
ad(x^j)\diamond_2 ad_2(x^k)$$
$$ + L_{ij,k}\diamond ad(x^k)\diamond_2ad_2(x^j). $$
Recall that 
$L_{ij,k}\diamond ad(x^k) = ad(L_{ij}).$ Hence we can rewrite the connection in
the new gauge as 

$$\partial_i + \lambda_i - P_i + (b-L)_{ik}\diamond ad(x^k) 
- (W - dP-L)_{ij})\diamond_2ad_2(x^j)$$
$$ - (c+dL)_{ijk}\diamond ad(x^j)\diamond_2 ad_2(x^k). $$

Thus we have a gauge transformation which transforms the three form $c$ 
to $c+dL$ and the two form $b-W$ to $b-W+dP$, while sending 
$\lambda_i$ to $\lambda_i-P_i$ and $(b+W)/2$ to $(b+W)/2  - L$.  This appears
to have a somewhat different field content from what we expect for type $IIA$
supergravity, where three form potentials such as $c$ arise. 
(See \cite{P2}{Chapter 12}). In subsequent work, we shall explore an alternate construction of a
quantum mechanical model for Ramond-Ramond fields. 

We may now consider various choices of representation: 
connections or gauge equivalence classes of connection, gauge equivalence
classes of $B$ field connection, etc.  One can iterate such constructions ad
nauseum.

\subsection{Non-abelian gauge groups}
It is natural to ask what happens if we allow our connections to take values in
a nonabelian group. So, we consider a "gauge transformation" 
$\nabla_{\frac{\partial}{\partial x^k}}\rightarrow 
\nabla_{\frac{\partial}{\partial x^k}} + \lambda_k.$ 
Here $\lambda_k$ is endomorphism valued.
$$ad(\nabla_{\frac{\partial}{\partial x^i}}
 - b_{ij}ad(x^j))\nabla_{\frac{\partial}{\partial x^k}} = F_{ik} + b_{ik},$$
but 
$ad(\nabla_{\frac{\partial}{\partial x^i}}
 - b_{ij}ad(x^j))(\nabla_{\frac{\partial}{\partial x^k}} + \lambda_k)
  = F_{ik} + \lambda_{k,i} + b_{ik}.$ 
Hence, as before, we must modify $ad(\nabla_{\frac{\partial}{\partial x^i}}
 - b_{ij}ad(x^j))$. We replace 
 $\nabla_{\frac{\partial}{\partial x^i}}$ by 
$\nabla_{\frac{\partial}{\partial x^i}} - \lambda_i$. We expect that we 
must replace 
$b$ by $b + d\lambda$, but now there is some choice as to which connection we use to
compute $d\lambda$, $d$ or $d+\lambda$. If we choose the average 
$d_{\lambda/2}:= d+\lambda/2$, then we get 

$$ad(\nabla_{\frac{\partial}{\partial x^i}} + \lambda_i
 - (b+d_{\lambda/2}\lambda)_{ij}ad(x^j))(\nabla_{\frac{\partial}{\partial x^k}} + \lambda_k)
  = F_{ik} + b_{ik},$$
  as desired. 
  
This extension of gerbes does not look suitable for string theory. In
particular, it is no longer clear how to decouple the B field from the
connection for closed strings. If we assume, however, that the $B$ field takes
values in the center, $C$, of the gauge group then we may pass to a closed
 string theory if we consider $C-$ gauge equivalence classes of pdos.

\section{Acknowledgements}
I would like to thank P. Aspinwall and W. Pardon for helpful discussions.


\begin{thebibliography}{99}


\bibitem{B1} T. Buscher, A symmetry of the string background field equations, 
Phys. Lett. B 194 (1987) p. 59.

\bibitem{B2} T. Buscher, Path integral derivation of quantum duality in
nonlinear sigma models, Phys. Lett. B 201 (1988) p. 466.

\bibitem{CJM} A. Carey, S. Johnson, and M. Murray, Holonomy on D branes, 
hep-th/0204199.

\bibitem{CK} D. Cox and S. Katz, {\em Mirror Symmetry and Algebraic Geometry},
 Providence, American Mathematical Society (1999).

\bibitem{Fed} B. Fedosov, {\em Deformation Quantization and Index Theory},
 Berlin, Akademie Verlag (1996).
 
\bibitem{FW} D. Freed and E. Witten, Anomalies in String Theory with D-Branes,
 hep-th/9907189.

\bibitem{H} N. Hitchin, Lectures on special Lagrangian submanifolds, 
{\em Winter School on Mirror Symmetry, Vector Bundles, and Lagrangian
Submanifolds}, 151-182, AMS/IP Stud Adv. Math. 23, Amer. Math. Soc., Providence,
RI, 2001 DG/9907034.

\bibitem{K} A. Kapustin, D-Branes in a topologically nontrivial B-field,
Adv. Theor. Math. Phys. 4 (2001) 127 hep-th/9909089.
   
\bibitem{KO} A. Kapustin and D. Orlov, Vertex Algebras, Mirror Symmetry, And
D-Branes: The Case of Complex Tori, hep-th/0010293.

\bibitem{OOY} H. Ooguri, Y. Oz, and Z. Yin, D-branes on Calabi-Yau Spaces and
Their Mirrors, hep-th/9606112, Nuc. Phys. B 477 (1996) pp. 407-430.
  
\bibitem{P} J. Polchinski, {\em String Theory Volume I},
 Cambridge, Cambridge University Press (1998).

\bibitem{P2} J. Polchinski, {\em String Theory Volume II},
 Cambridge, Cambridge University Press (1998).
  
\bibitem{VS} V. Schomerus, D-branes and Deformation Quantization,
geometry, hep-th/9903205, JHEP 9906 (1999) 030.

\bibitem{SW} N. Seiberg and E. Witten, String theory and noncommutative
geometry, hep-th/9908142, JHEP 9909 (1999) 032.

\bibitem{SYZ} A. Strominger, E. Zaslow, and S.-T. Yau, Mirror Symmetry is T
duality, hep-th/9606040, Nuc. Phys. B 479 (1996) pp. 243-259.

\end{thebibliography}
\end{document}